\documentclass[10pt,a4paper]{article}

\usepackage[utf8]{inputenc}

\usepackage{cite}
\usepackage[justification=centering]{caption}

\usepackage{nameref,hyperref}

\usepackage{microtype}
\DisableLigatures[f]{encoding = *, family = * }

\usepackage[aboveskip=1pt,labelfont=bf,labelsep=period,singlelinecheck=off]{caption}

\makeatletter
\renewcommand{\@biblabel}[1]{\quad#1.}
\makeatother

\usepackage[margin=2.5cm]{geometry}

\usepackage{color}

\definecolor{Gray}{gray}{.25}

\usepackage{graphicx}

\usepackage{amsmath}
\usepackage{amsfonts}
\usepackage{amssymb}

\newcommand{\De}[3][]{\frac{\partial^{#1}#2}{\partial {#3}^{#1}}}
\newcommand{\de}[3][]{\frac{\mathrm{d}^{#1}#2}{\mathrm{d}{#3}^{#1}}}
\renewcommand{\d}{\,\mathrm{d}}

\renewcommand{\eqref}[1]{(\ref{eq:#1})}
\newcommand{\figref}[1]{\ref{fig:#1}}
\newcommand{\tabref}[1]{\ref{tab:#1}}

\catcode`\"=\active
\def"#1"{``#1''}

\let\at=@
\catcode`\@=\active
\def@#1{\ifmmode\boldsymbol{#1}\else\at#1\fi}

\usepackage{ifthen}
\usepackage{xspace}

\newcommand{\eq}[2][]{%
    \ifthenelse{\equal{#1}{}}{%
        \begin{equation*}
            #2%
        \end{equation*}%
    }{%
        \begin{equation}\label{eq:#1}%
            #2%
        \end{equation}%
    }%
}
\newcommand{\meq}[2][]{%
    \ifthenelse{\equal{#1}{}}{%
        \begin{equation*}%
            \begin{split}%
                #2%
            \end{split}%
        \end{equation*}%
    }{%
        \begin{equation}\label{eq:#1}%
            \begin{split}%
                #2%
            \end{split}%
        \end{equation}%
    }%
}

\newcommand{\lfig}[1]{\label{fig:#1}}
\newcommand{\ltab}[1]{\label{tab:#1}}

\begin{document}

% title goes here:
\title{An Elementary Microscopic Model of Sympatric Speciation}
% authors go here:
\author{
Franco Bagnoli  and 
Tommaso Matteuzzi
\\
Dept. Physics and Astronomy, and CSDC, University of Florence,\\
Via Sansone 1, I-50019, Sesto Fiorentino, Italy. Also  INFN, Sez. Firenze.
\\
\{franco.bagnoli,tommaso.matteuzzi\}@unifi.it
}

\maketitle

\section*{Abstract}

Using as a narrative theme the example of Darwin's finches, a microscopic agent-based model is introduces to study sympatric speciation as a result of competition for resources in the same ecological niche. Varying competition among individuals and resource distribution, the model exhibits some of the main features of evolutionary branching processing. The model can be extended to include spatial effects, different genetic loci, sexual mating and recombination, etc. 
% now start line numbers
%\linenumbers

% the * after section prevents numbering
\section{Introduction}

Charles Darwin published \textit{On the origin of the species} in 1859. However, an early idea of the theory of evolution was already in his mind many years before. In  \textit{The voyage of the Beagle} \cite{Voyage}, published in 1845, he comments on gradation in the size of the beaks in different species of finches of Galapàgos islands (Fig.~\figref{finches}): \textit{“Seeing this gradation and diversity of structure in one small, intimately related group of birds, one might really fancy that from an original paucity of birds in this archipelago, one species had been taken and modified for different ends”}

\begin{figure}[t]
\includegraphics[width=12cm]{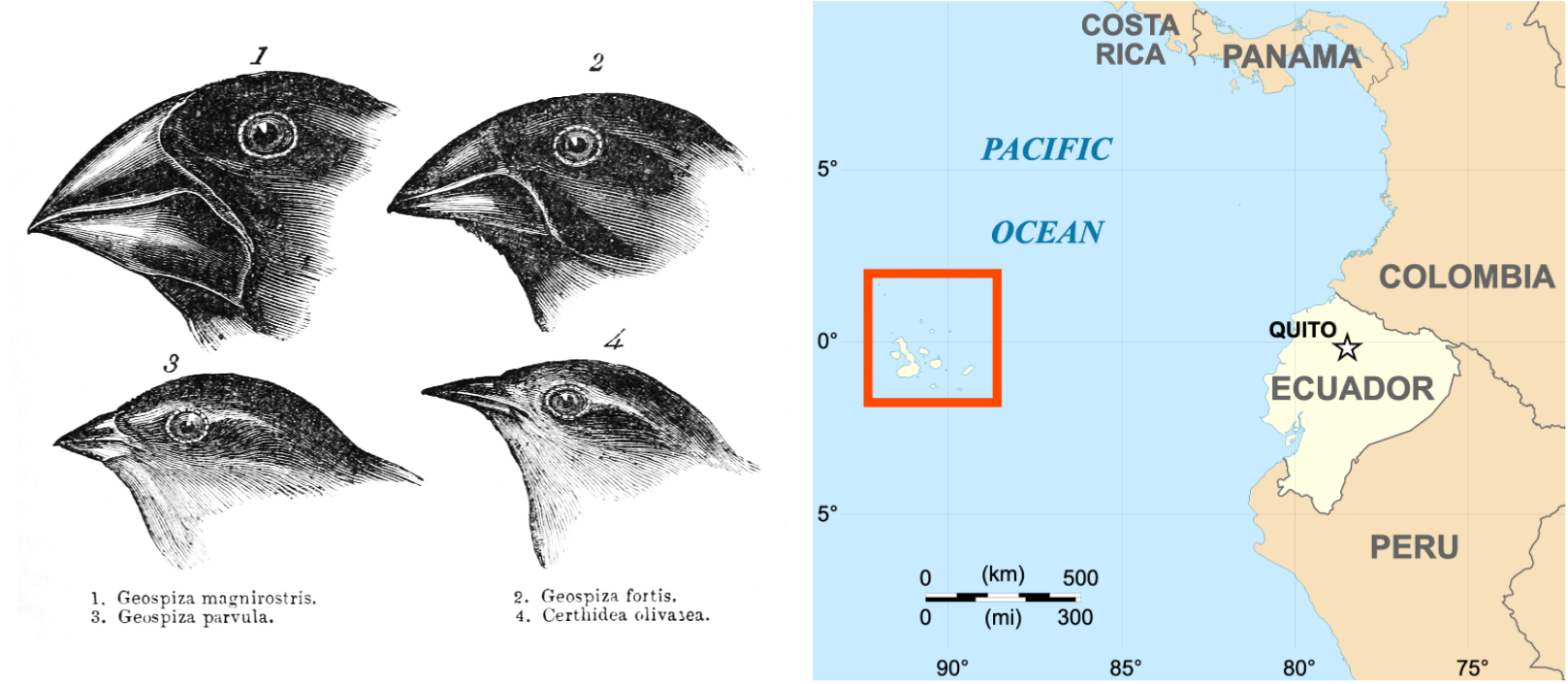}\\[2mm]
\centering
\caption{\textbf{Darwin's finches.} (left) Four of the species observed by Darwin in the Galapagos island during its voyage on the Beagle. The drawings are due to the zoologist J. Gould who first recognized the birds as a new peculiar group of finches. (right) Galapagos islands location in the Pacific Ocean.}
\lfig{finches}
\end{figure}

Darwin was a keen observer as these species of birds, now known as Darwin's finches, are one of the most strikingly example of evolution at work. 

Darwin's finches \cite{Steinheimer2004, Lamichhaney2015} are a group of several related birds species, which differ in beak size and shape, originated from a series of evolutionary branching events, i.e. speciations, in the Galapagos islands. 

An evolutionary branching occurs when an existing population of organisms, all with the same phenotype, splits in two coexisting populations with different phenotypic traits. 

Speciation can be \textit{allopatric} or \textit{sympatric}. Allopatric means “in different places” and is due to the presence of a geography barrier between two populations. One of the possible reasons for differentiation is that the two populations undergo different selective pressures (i.e., different environments), while the exchange of genes is forbidden by the barrier. 

Another divergence source is due to genetic drift. In finite populations, mutations can fix if neutral and also if slightly deleterious, the so-called neutral evolution \cite{Kimura1968,Kimura1969}. This is the basis of the ``molecular clock'', i.e., the correspondence between genetic distance and time since last common ancestor. More complex genetic rearrangements, like chromosome splitting or fusion, can forbid inter-fertility and thus lead to a complete separation of species, as happened with humans and chimpanzees~\cite{Ayala2005}. 

The concept of species is generally related to sexual reproduction: individuals belongs to different species if they cannot breed (or if they generate infertile offsprings), or if they, although being inter-fertile, do not generally breed in normal situations (for example periodical cicadas~\cite{Tanaka2009}). 

However, one can define species  also for asexually-reproductive populations \cite{WilliamBirky2009, Cohan2002}, in this case the difference among species if due to phenotypic or genotypic differences, in the absence of intermediate cases. 

For both sexual and asexual populations, the usual assumption is that different species occupy different niches, and that speciation occurs after a modification of the environment. In particular, for bacteria and other unicellular organisms, it is assumed that mutations allow to quickly explore the possibilities of the environment (for instance, possibility of parasitism), so that all possible niches are quickly discovered and occupied. 

However, there are documented cases of sympatric (``in the same place'') speciation. In this case, populations experience the same selective pressures and gene exchange is allowed. 

Probably, the first example of sympatric speciation is that of Darwin finches, which is important for at least two reasons: on one hand, they are an example of sympatric \textit{adaptive radiation}, i.e. starting from a single ancestor in a new environment, a cascade of speciation events gave rise, in a (geologically) short time, to a variety of new species all sharing the same environment~\cite{Voyage}.

On the other hand, the radiation event of Darwin finches is relatively young (Galapagos formed only few millions of years ago) so that phenotypic differences between species are small (i.e., they differ mainly in the beaks, while other features, e.g. feathering, remain similar) and their direct ecological basis can be investigated and modeled. 

There are other well documented examples of sympatric speciation, for instance that of cichlid fishes in volcanic lakes \cite{Barluenga2006,Kocher2004}. Also in this case, it is assumed that a single-species colonization of the rather uniform environment has give origin to a well differentiated communities of non-interfertile species.

Finally, also the formation of different strains of infective agents, like bacteria or viruses, within a single individual or a host population can be seen as the prototype of sympatric speciation. 

The main question is whether these speciation events are due to the occupation of pre-existing niches, or if they are caused by competition in a rather uniform environment. 

We shall adopt here a description based on Darwin finches for historical reasons and also because this image is more evocative. So we shall speak of a phenotypic trait like the size of beaks of finches, and of resource distribution like the size distribution of available seeds. Competition arises because finches with a given beak size are assumed to be able to profitably feed on seeds of a certain size range, since they cannot feed on those too much bigger than their beak size, and feeding on those too small will not furnish enough energy (assuming that the beak size and the body size are correlated). 

The distribution of seeds in the absence of birds gives the ``static'' fitness landscape. Assuming that this is smooth and single-peaked (for instance a Gaussian), there is only one ``niche''. Therefore, it can be naively assumed that the resulting population will occupy that niche, possibly with a broad distribution. 

\begin{figure}[t]
    \centering
    \includegraphics[width=0.5\linewidth]{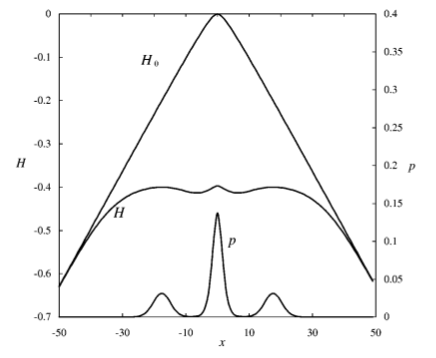}
    \caption{Static fitness $H_0$ and dynamic fitness $H$, due to the presence of a three-peaked distribution $p$. Note that the peaks correspond to the maxima of the dynamic fitness. Image from Ref.~\cite{Bagnoli1997}.}
    \label{fig:Bagnoli1997}
\end{figure}

There are some simple models showing how sympatric speciation can arise due to competition \cite{Bagnoli1997,Dieckmann1997,Dieckmann_nature,Bagnoli2005,Bagnoli2005a}. In Ref.~\cite{Bagnoli1997} speciation is observed in asexual populations as due to competition. After a colonization event, with a non-optimal phenotype, population tends to vary due to mutations and occupy the niche. However, at this point, the presence of birds consuming resources, changes the shape of fitness. If the range of competition is small and the peak of the static fitness is rather broad, other well-separated peaks in the bird distribution can appear (Fig.~\ref{fig:Bagnoli1997}). The peaks correspond to maxima of the dynamic fitness (i.e., of seed availability in the presence of feeding birds), which corresponds to the Gause principle \cite{Gause1932,Pocheville2015}. Actually, the conditions for coexistence can be obtained in a self-consistent way by imposing this condition, as shown in Section~\ref{sec:methods}. 

Speciation is more difficult to obtain in the case of sexual reproduction, or, better, of recombination. Gene exchange in the population favor intermediate phenotypes. One possibility is that of assuming assortative mating, i.e., mating only within individuals carrying certain phenotypic characteristics~\cite{Bagnoli2005,Bagnoli2005a}.

A coevolutionary approach to sympatric speciation, including the case of sexual recombination, was developed by Dieckman and Doebeli~\cite{Dieckmann_nature}. In this model authors re-obtained the previous results for asexual speciation, but also developed a sophisticated model for examining the occurrence of the same phenomenon in the sexual case.  They studied the case of assortative mating due to another genetic character. Individuals therefore carry a set of genes determining their phenotypic ``ecological'' character, i.e., beak size, plus another set of genes determining their mating preferences with individuals with a certain beak size. 

It may happen that individual with small beak prefer mating with individual with similar, or with different beak size, according with the ``preference'' genes. The result is that, if the static fitness (abundance of seeds) is broad enough, population tends to split in two ``species'', formed by individuals that mate assortatively with other individuals in the same population, while those with mixed phenotype-mating preferences disappear. This splitting causes that no species occupy the natural ``niche'' of most abundant resource, corresponding to an intermediate phenotype. Actually, as shown in Section ~\ref{sec:methods}, it could happen that the original species stays in the maximum-fitness location, but this would imply that the surrounding newly formed quasispecies are symmetrically placed. In general, only one of them appears, thus “pushing” the original species away from the maximum of fitness.

The model may also include the effect of ornaments. In this case individuals are assumed to carry a third set of genes, specifying for an ornament (color of plumage, length of tail, etc.) with no fitness effect, and the mating preferences determined by the second set of genes now select for ornaments, not for the ecological character. 

Also in this case, but with more stringent conditions, speciation can occur. The population again splits in two species, exhibiting both different ornaments but also different beak size. The linkage disequilibrium in the ecological character is now due to genetic drift: population splits due to coupling between ornament and mating preferences (individuals carrying a type of ornaments but preferring mating with individuals with opposite ornament disappear). In this splitting, random assortativity of the ecological character occur, and this brings to a linkage disequilibrium also in the (uncorrelated) ecological character. 

However, these models are of mean-field type, assuming that the population is well-mixed, i.e., they do not include spatial effects. Some of them are phenotypic, i.e., consider only the abundance of some phenotypic character, disregarding the genotypic aspect, and in general they do not directly include the presence of resources, but only consider their indirect effect through an interaction term, as illustrated in Section  ~\ref{sec:methods}. To be specific, instead of considering the coupling between the abundance of seeds (that depends on their production and consumption by birds) and birds (whose survival depends on the abundance of seeds), like in a prey-predator system, one assumes that the abundance of seeds immediately reached its asymptotic value, so that the presence of a phenotype that feeds on a certain range of seed sizes, is felt by birds with similar phenotype as a competition term that depends on the abundance of the two strains. 

The goal of this paper is that of examining the necessary components for designing a real microscopic (spatial) and genotypic model, in which both birds and seeds time evolution are considered, in a real agent-based approach. We think that such a model can be more convincing for didactic purposes. We limit our exposition here to an asexual model, which is shown being already quite complex. The sexual case is deferred to a subsequent publication. 

\section{A Simple Phenotypic Model}\label{sec:methods} 

In order to smoothly introduce the subject, and also to revise the advantages and drawbacks of classical population dynamics approach, let us introduce a simple phenotypic model, similar to that introduced in Ref.~\cite{Bagnoli1997}.

Let us suppose that $x$ denotes the phenotypic space, $0 \le x \le 1$ ($x$ is considered a continuous quantity) and $n(x,t)$ is the number of animals with phenotype $x$ at time $t$. The phenotype can be for instance proportional to the beak size.

The total number of animals at time $t$ is $N(t) =\int  n(x,t)\,\mathrm{d}x$, and the probability distribution of animals is $p(x,t)=n(x,t)/N(t)$.

The resources, for instance seeds, are produced at a rate $W_0 (x)$ and are destroyed at a rate $q$, so that, in the absence of animals, their instantaneous abundance $W(x,t)$ is given by
\eq{
    \De{W}{t}=W_0 (x)-qW(x,t).
}
The stationary distribution of resources ($\partial W/\partial t= 0$) is therefore $\overline{W}(x) = W_0 (x)/q$.

Let us now suppose that there is an animal distribution $n(x,t)$ and that animals with phenotype x can feed on seeds of size $x$, so we use the same phenotypic index for beak and seed size. We also assume that the energy income of birds are proportional only to the number of eaten seeds, and not on their size, since birds with larger beaks should have comparable larger body mass, and therefore feeding on larger seeds imply eating the same number of seeds as having smaller beak and body, and feeding on smaller seeds.

If every phenotype feeds only on the seeds of corresponding size, we have
\eq{
    \De{W}{t}=W_0 (x)-qW(x,t)-\alpha n(x,t)W(x,t),
}
where $\alpha$ determines the fraction of seeds consumed and may depends on $x$.

We now assume that the dynamics of seeds is much faster that that of population, we can consider that seeds distribution is at any time neat to the asymptotic one, and therefore
\eq{
    \overline{W}(x,t)=\frac{W_0(x)}{q+\alpha n(x,t) }.
}

Since we are interested in the effects of competition, we assume that $q$ is quite large with respect to the population feeding on this resource, so that there is a shortage of food. Therefore we can write
\eq{
    \overline{W}(x,t)\simeq \frac{W_0(x)}{q }\left(1-\frac{\alpha}{q} n(x,t)\right)=H_0 (x)-Kn(x,t).
}

If however now we consider that birds with phenotype $x$ can also feed on seeds of size $y$, with an efficiency $K(x - y)$, we have
\eq{
    \overline{W}(x,t)=H_0 (x)-\int  K(x-y)n(y,t)\d y,
}
and this introduces the competition for resources.

The evolution equation of the population is
\eq{
    \De{n}{t}=(A(x,t)-\gamma )n(x,t),
}
where $A$ is the birth rate and $\gamma $ the death rate, which is supposed to be the same for all phenotypes. The total population equation $(N(t)=\int n(x,t)dx)$ is
\eq{
    \de{N}{t}=\int A(x,t)n(x,t)\d x-\gamma N
}
We now switch to probability distribution $p(x,t) = n(x,t)/N(t)$. Taking the time derivative of $p$
(with implies the time derivative of $n$ and $N$), and with a bit of algebra we get
\eq[p]{
    \De{p}{t}=(A(x,t)-\langle A\rangle(t))p(x,t)
}
where $\langle A\rangle(t)=\int A(x,t)p(x,t)\d x$.

Notice that the equation for the probability distribution alone is not sufficient, it might happen that $N$ vanishes due to insufficient resources with respect to the death rate (which does not appear in the equation for $p$).

The meaning of Eq.\eqref{p} is clear: the phenotypes with birth rate less than average tend to disappear, the other to increase. However, the fitness $A$ in general depends on the population. If
we consider $A$ proportional to $W$, we get
\eq{
\De{p}{t}=(H_0 (x)-\int K(x-y)p(y,t)\d y-\langle A\rangle(t))p(x,t).
}
The term $\langle A\rangle(t) = \phi (t)$ is only needed to keep p normalized, in a numerical simulation it can be avoided, simply renormalizing p at each time step (so that $\int p(x,t)dx = 1$).

We also need to add a mechanism to populate new phenotypes, in case they are not present at beginning. This can be done via mutations, for instance adding a term like $\mu\nabla ^2 p$. In general, the effects of mutations are only that of insuring a unique asymptotic distribution, and that of determining the time scale, although there can be influences on extinction and quasispecies distributions, in the case of very rough fitness landscapes~\cite{Bagnoli1998}. In any case, the effect of mutations is that of broadening the peaks of the distribution.

Instead of examining all possibilities, let us concentrate on a situation that we want to investigate using the microscopic model illustrated in the following, i.e., that of a speciation event induced by competition. We can assume that the "static fitness" $H_0 (x)$ is smooth and single peaked, at phenotype $x_0$. We also assume that the mutation rate is vanishing, so that the asymptotic distribution is formed by very narrow peaks. We may consider the possible presence of a pair of symmetric species at phenotypes $x_(-1)  = x_0  - \delta x$ and $x_1= x_0+ \delta x$. We shall denote with $p^{0)} (t) = p(x_0,t)$, $p^{(1)} (t) = p(x_(-1),t) = p(x_1,t)$ and $H_0^{(0)}=H_0 (x_0 )$,$ H_0^{(1)}=H_0 (x_1 )  =H_0 (x_(-1) ))$. We also denote $K = K(\delta x)$.

Disregarding mutations and the competition among the two satellite species $p^{(1)}$, we have
\eq[pp]{
    \begin{cases}
         \displaystyle \de{p^{(0) )}}{t}=\left(H_0^{(0)}-2K p^{(1)}-\phi \right) p^{(0)},\\[4mm]
        \displaystyle \de{p^{(1)}}{t}=\left(H_0^{(1)}-K p^{(0)} -\phi \right) p^{(1)},
    \end{cases}
}
with $p^{(0)}  + 2p^{(1)}  = 1$. Imposing this condition (so that its time derivative is null), we get
\eq{
    \phi (t)=H_0^{(0)} p^{(0)}+2H_0^{(1)} p^((1) )-4Kp^{(0)} p^{(1)}.
}

The system of Eq.~\eqref{pp} can have asymptotic solutions $p^{(0)}=1$ and $p^{(1)}=0$; $p^{(0)}=0$ and $p^{(1)}=1/2$, or $p^{(0)}\neq 1$ and $p^{(1)}\neq 0$.
    
In the third case, one can easily check that the asymptotic solution for $p^{(0)}$,
\eq[HH]{
    p^{(0)}=\frac{1}{2} \left(1-\frac{H_0^{(0))}H_0^{(1)}}{K}\right),
}
is acceptable ($p^{(0)}>0$) for $H_0^{(0)}-H_0^{(1)}<K$. This is the condition for coexistence, and by inserting
the explicit expression for the static fitness and the competition kernel one can get the
“equilibrium” distance $\delta x$ and the quasispecies sizes. However, this solution is unstable, so that starting below the value given by Eq.\eqref{HH}, $p^{(0)}$ goes to zero and above it $p^{(0)}$ goes to one.
If the asymptotic value of Eq.\eqref{HH}) goes below zero, i.e., for $H_0^{(0)}-H_0^{(1)}>K$, only the “central” quasispecies is present, if it goes above one only the “peripheral” ones survive, but this cannot happen if $H_0^{(0)}>H_0^{(1)}$.

More sophisticated population dynamics can be proposed, but we think that this first exercise is sufficient to set up questions that are to be investigated by means of a microscopic model:
\begin{itemize}
    \item	Is stable coexistence possible?
    \item	Is the inclusion of genotypes (and the consequent degeneracy of phenotypes) affecting the scenario?
    \item	Does the explicit dynamics of seeds change the system?
    \item	What about the influence of spatial dynamics?
\end{itemize}
Only some of these questions will be answered in this first paper.

\section{The microscopic approach} 
We are interested in investigating a spatial system, for instance modeling the dispersion of birds on an island.

\subsection{Modeling Scenario}

An island is approximated by a set of patches, arranged in a square grid. It is populated by many plant species, each species producing seeds of different sizes, with a given distribution. At the beginning, there are no animals eating seeds so they accumulate at the ground and degrades. Plant species are distributed homogeneously in space.

At a certain time, a particular bird species arrives in the island. The incoming species own a certain phenotype, i.e., beak size, which determines its ability of eating seeds within a given size range. If there are enough seeds of the right size, birds start to reproduce and population grows. 

Since reproduction is affected by mutations, some offspring will show a different beak size (smaller of grater) and will be able to eat seeds of other dimensions. Our research topic is how  the distribution of phenotypes evolves in time. 

\subsection{The Model}

We simulated the above scenario building the following agent-based microscopic model (Fig.~\figref{lattice}).

The spatial setting of  our ecosystem (i.e., the island) is a two-dimensional $N$ by $N$ square-grid. In each patch (i.e., grid square) there is same food. According to our scenario, we assume the food to be plant seeds but it could be any other nutrient, for example, insects (for birds), algae, plankton ( for fish) or some chemicals (for bacteria). 

Food is characterized by a continuous parameter, $x$, taking values in the interval $[0,1]$. In our case, it represents an appropriately normalized seeds dimension. 

\paragraph{Seeding.}
Seeds of size $x$ are created  at a rate $H_0(x)$ and destroyed at a constant rate $d$. The function $H_0$ can be thought as the \textit{static fitness}, proportional to the phenotypic carrying capacity. The actual availability of seeds, $H(x)$, depends on the presence of phenotypes feeding on seeds of size $x$.

In the following, we assume $H_0(x)$ to be proportional to a Gaussian $H_0(x) =rN(a,b)$, where $a$ is the average and $b$ the standard deviation, normalized over the interval $[0,1]$.

Seeds have a fixed content of energy per feeding individual $e$, regardless of their size, since we assume that individuals with larger beak (feeding on bigger seeds) also have proportionally larger body size, so they need more incoming energy than smaller birds feeding on smaller seeds.

\paragraph{Colonization. }
After a transient time $(t_0)$, during which seed distribution reaches a steady-state, some birds, all with the same phenotype, are created in a given position on the grid.

\begin{figure}[t]
\includegraphics[width=10cm]{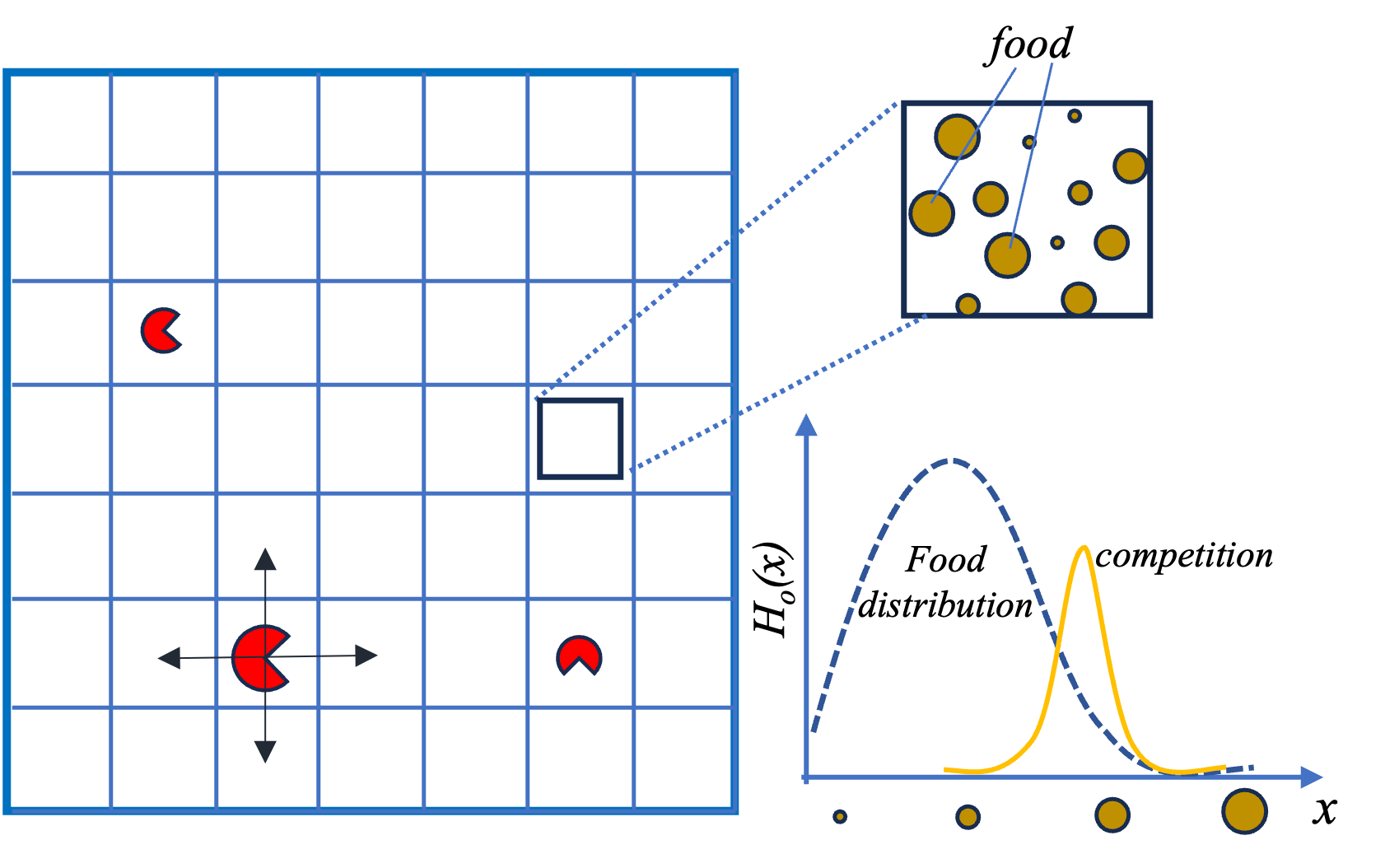}
\centering
\caption{\textbf{Scheme of the microscopic model}. Seeds (yellow circle) of size $x$ (in the interval $(0,1)$) are distributed in each cell according to the distribution $H(x)$. Birds (red pacmen) moves on the grid searching for a seed of the right size, affecting the resulting seed abundance.}
\lfig{lattice}
\end{figure}

\begin{table}[t]
    \caption{Microscopic model parameters.}
    \ltab{para}
    \centering
    \begin{tabular}{||c c c c||} 
        \hline
        Parameter & - -  & Range & Simulations\\ [0.5ex] 
        \hline\hline
        Seeds distribution mean  & $a$ & $(0,1)$ & variable \\ 
        Seeds distribution std  & $b$ & $(0,1)$ & variable \\
        Seeds deposition rate & $r$ & $(0,1)$ & 0.2  \\
        Seeds depletion rate & $d$ & $(0,1)$ & 0.1  \\
        Initial phenotype & $x_0$ & $(0,1)$ & 0.6 \\
        Initial number of birds & $N$ & - & 10 \\
        Energy per seed & $e$ & - & 0.16 \\
        Reproduction threshold & $E_{th}$ & - & 2 \\
        Bird genotype length (bit) &  $L$  & $2 - 64$ & $64$  \\  
        Mutation probability & $\mu$ & (0,1) & 0.01  \\
        Energy consumption per time step & $e^-$ & - & 0.15  \\ 
        Competition range & $c$ & (0,1) & variable   \\ [1ex]
        \hline
    \end{tabular}
\end{table}

\begin{figure}[t]
    \includegraphics[width=10cm]{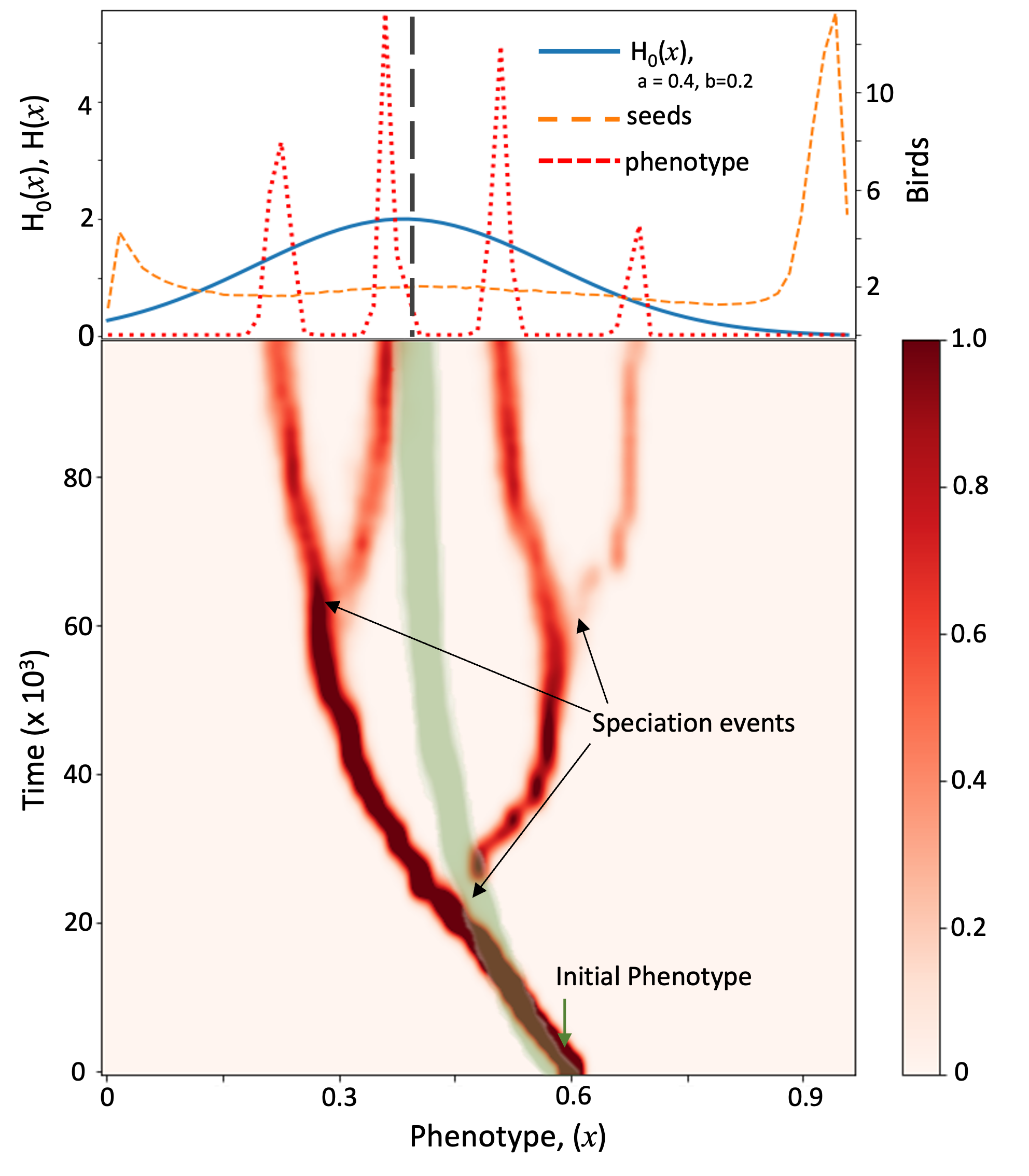}
    \centering
    \caption{\textbf{Evolution of phenotypic distribution in time}. Static fitness (top panel, blue line) is slightly asymmetric $a=0.4, b = 0.2$. The initial phenotype population (i.e. colonizing species, $x_0=0.6, N = 10$) evolves toward the fitness maximum and undergoes a series of evolutionary branching events (red trajectory) if the competition range is fixed to $c=0.1$. In this case, the steady-state phenotypic distribution (at $T = 10^5$) shows four well separated peaks (top panel, red dotted line). Otherwise, if $c \geq b $, the population stabilizes around the fitness maximum (green trajectory, $c=0.2$).}
    \lfig{evo}
\end{figure}

\paragraph{Movement.} Birds behave like walkers on the grid and change position at every time step. This movement results in an almost homogeneous distribution of phenotypes and therefore seed availability. In future experiment we plan to limit bird movements so to explore the spatial consequences.

\paragraph{Grazing. }

Each bird, $i$, is characterized by an energy level ($E_i$), a genotype ($G_i$) and a phenotype ($P_i$), the phenotype being the beak size.

The genotype is modeled by a string of $L$ bits. Given a genotype, $G=(g_1, \dots, g_L)$, $g_i=0,1$, the associated phenotype is given by
\begin{equation}
    P(G) = \frac{1}{L}\sum_i g_i,
\end{equation}
meaning that the genotype-phenotype mapping is additive. The phenotype (i.e., beak size) takes values in the same interval of seed dimensions, $[0,1]$. 

A bird can, in principle, eat seeds of every dimension, but has a grater ability to eat seeds whose size is more similar to its phenotype.

To mimic this, in the model when a bird of beak size $x_0$, finds a seed of dimension $x$, it eats the seed with a probability modeled as a Gaussian $\mathcal{N}(x_0,c)$, centered on $x_0$ and with standard deviation  $c$. 

We called the parameter $c$ the \textit{competition range}, since birds with similar beak size compete mostly for seed in that range.

At every time step, a bird make some attempts to eat a seed, while its energy decreases of a fixed quantity $e^-$. If it finds food, its level of energy increase of $e$. Birds die when they run out of energy (i.e, they do not find enough seeds) while, if their energy exceed a threshold $E_{th}$, they reproduce. 

\paragraph{Reproduction. }
Reproduction is asexual. When an existing bird has $E>E_{th}$ its energy is halved and a new bird is created with an energy level $E/2$. 

The genotype of the new bird is a copy of the parent's genotype with mutations i.e., flipping of one ore more bits. The mutation rate, $\mu$, is the probability of flipping one  bit of the genome during the reproduction.

As a consequence, the child can have a beak size (phenotype) which differs from its parent and therefore will be able to eat seed in a different size range. 

\paragraph{Competition. }
Birds in the population compete for seeds. As said before, the range of competition between different phenotypes is tuned by the parameter $c$, which can be roughly seen as the range of seeds size eaten (with reasonable probability) by a given phenotype. ``Specialized'' species have a small competition range while species with larger $c$ are ``generalist''. In principle, one could allow $c$ to depend on the genotype and thus to evolve, but this requires some careful considerations. Generalistic species can feed on a larger rage of food, but this capacity should come at the price of being less efficient than specialized species, in order to allow both ``strategies'' to thrive in the right conditions. One expects ``specialized'' strategies to emerge in static environment, and generalistic ones in time-varying conditions. 

In the present model, $c$ is fixed at the beginning of the simulation and do not evolve in time, i.e. each individual and each species has the same degree of specialization. 

\subsection{Simulations}

We run simulations with a $80 \times 80$ square-grid, changing competition range $c$ and seed distribution parameters ($a,b$). 

The other parameters of the model where held fixed as they do not significantly affect the speciation behavior. 

In order to sustain reproduction, the energy per seed $e$ has to be grater than the energy consumption $e^-$. The ratio of the rates of seed deposition $r$ and depletion $d$ influences the asymptotic number of seeds and birds but do not influences the shape of the final distribution, unless in the case of extinction. 

The numerical values of parameters used for simulations are reported in table \tabref{para}.

\section{Results and Discussion}

As reported in the Introduction, several models suggest that sympatric speciation is the outcome of competition for resources without the need of pre-existing ecological niches. Many of this model are mean-field, do not take into account spatiality or do not include genotypes. We tested such predictions in a real microscopic spatial model which include genotypes, even though minimal.

Since we restrict our model to asexual reproduction, for us a speciation event, or evolutionary branching, is defined as the appearance of multimodality in the distribution of phenotypes. That is, the emergence, from an initial uni-modal distribution, of well-defined phenotypic clusters without intermediate cases. 

The static fitness, $H_0(x)$, is always assumed to be unimodal to accommodate the presence of only one "niche" in our environment.   

The typical evolutionary behavior of the model is illustrated in Fig.~\figref{evo}. If the initial phenotype is far from the maximum of the the static fitness (black dashed line), mutations and selection drive the population to that maximum. 

At this point, the distribution can be stable and unimodal, with a maximum in correspondence of the fitness maximum (green trajectory), or can split in two or more phenotypic sub-population, with the disappearance of the ``fittest type''. In this case, the steady-state distribution shows sharp peaks (top panel, red dotted line) highlighting the emergence of discrete population clusters, i.e, species.

Here, the critical role is played by the interplay between resources variability, summarized by the spreading of the seed-size distribution $b$ and intra-species competition $c$, which encodes birds specialization in resource usage. The smaller $c$, the higher the specialization.

The bifurcation diagram (Fig.~\figref{tree}, left panel), obtained by varying  the competition range while keeping fixed the distribution of resources, shows a first transition to bimodality when $c < b$. In other words, when the range of resources accessible by a species is smaller than the spreading of the resources, intra-species competition induces the splitting of the population. 

Decreasing $c$, the number of species at steady-state increases to  cover the entire range of available resources (Fig.~\figref{tree}, inset of left panel). 

\begin{figure}[t]
\includegraphics[width=14cm]{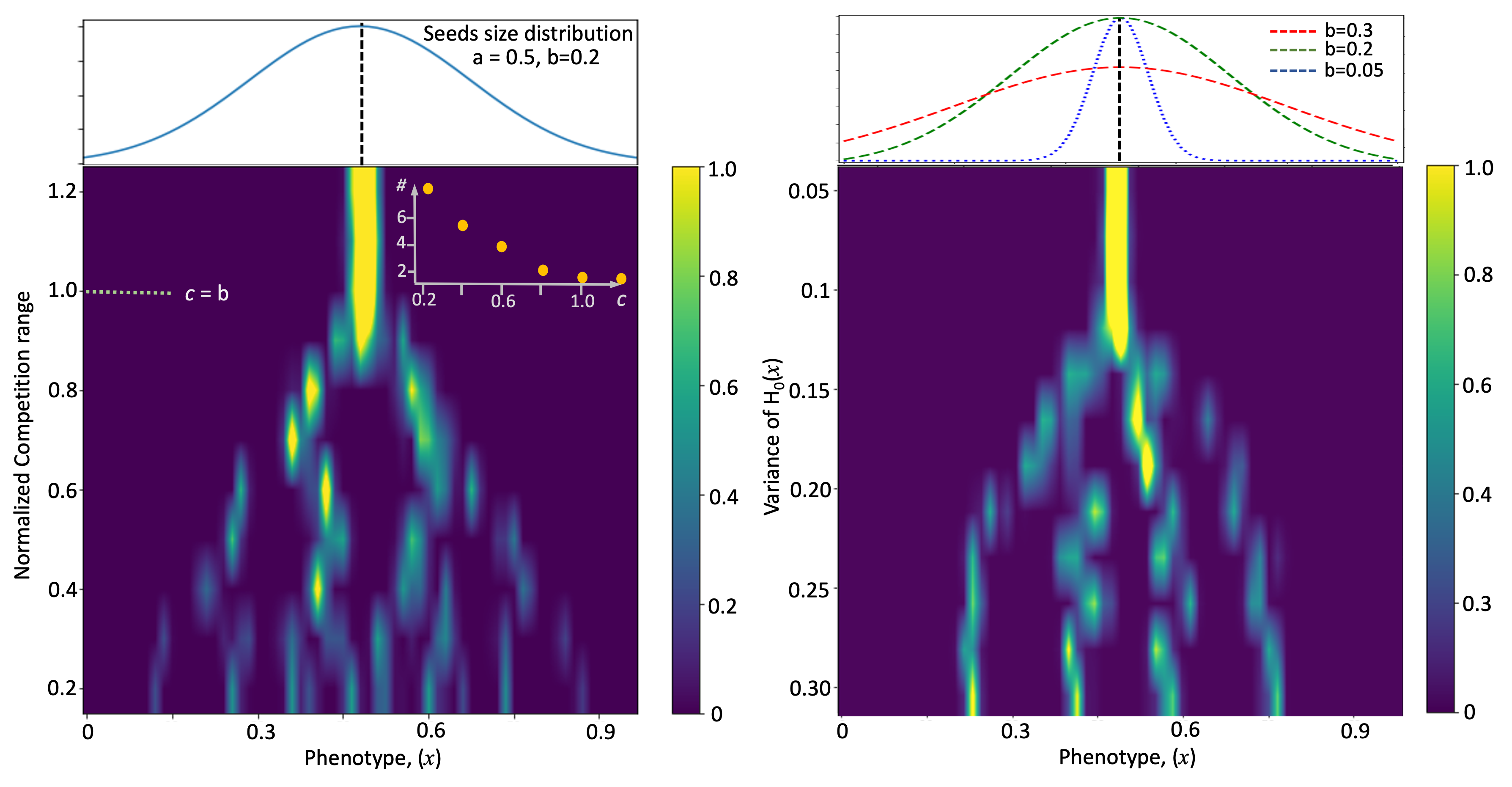}
\centering
\caption{\textbf{Species Formation from competition for resources.} Well defined phenotypic clusters (species) appears  when the competition range $c$ exceeds the standard deviation of the distribution of seeds, $b$. (Left) Phenotypes distribution (at steady-state, $T=10^5$) as a function of the normalized competition range $c/b$, where $b=0.2$, is held fixed. Inset: mean number of species at steady-state vs normalized $c$. (Right) Asymptotic phenotypes distribution ($T=10^5$) varying $b$ keeping $c=0.1$ fixed.}
\lfig{tree}
\end{figure}

A similar bifurcation diagram results keeping fixed the competition range and varying $b$ (Fig.~\ref{fig4}, right panel). Also in this case, the first transition occurs under the condition  $c < b$. 

These results are robust under changes of model parameters (system size, population density and energy consumption) and are in good agreement with previously cited models of speciation in absence of sexual reproduction \cite{Bagnoli2005a,Dieckmann_nature}.

It is interesting to note that typically, during a speciation event, population undergoes strong fluctuations and in one of the two branches the number of individuals can be very small. This can cause the fixation of random genotypic differences between the two formed species increasing their divergence in a way that resembles the founder effect \cite{Mayr1963,Jason2012}.

In Ref.~\cite{Dieckmann_nature}, a similar mechanism has been suggested to promote the linkage disequilibrium between sexual and ecological traits, leading to proper speciation, i.e. mating only within the species. Since our model incorporates genomes, a deeper investigation of genetic drift associated to speciation events will be the subject of a future work along with the extension of the model to sexual reproduction.

%possible extension of the model: 1)  species  

%\begin{figure}[h]
%\includegraphics[width=10cm]{varying_competition_bifurcation}
%\centering
%\caption{\textbf{Bifurcation varying competition range}}
%\label{fig3}
%\end{figure}

\section{Conclusions}

We illustrated a microscopic agent-based model that can be used for didactic purposes to illustrate sympatric speciation as a result of competition for resources in the same ecological niche. 

We investigated the effect of competition on the speciation phenomena.

For smooth fitness landscapes, and in the presence of competition, mutation only affects time scales and the broadening of peaks. In the present simulations, we moreover used a large diffusion rate so that also the spatial distribution does not influence the results. We checked that other parameters do not significantly affect the results.

The first step, which is what is carried on in this paper, is the reproduce the main characteristics of a standard phenotypic model using a genotypic, spatially-extended (i.e., agent-based) one.
Therefore, we imposed quite a fast diffusion of agent so that spatial effects are not present.

As said, we plan to extend the model including several other aspects, like sexual mating, recombination and the effects of resource fluctuations, both in space and in time, to understand how they affect speciation, extinctions and ecosystem resilience. 

We also would like to develop a friendly interface (at present in the form of a menu), although we also need to devote several efforts for its efficient implementation, since large populations and several time steps are needed to show the relevant effects.

We think our model could a valuable didactic tool to teach evolution, adaptability and ecology.

\section*{Code availability}
The simulation code (in C) is available upon requests to authors.

%\nolinenumbers

%This is where your bibliography is generated. Make sure that your .bib file is actually called library.bib
\bibliography{library}

%This defines the bibliographies style. Search online for a list of available styles.
\bibliographystyle{abbrv}

\end{document}